\documentclass[english, a4paper]{article}
\RequirePackage{geometry}
\usepackage[utf8]{inputenc}
\usepackage{natbib}
\usepackage{amsmath, amsfonts, bbm}
\usepackage{tikz-uml}

\usepackage{xspace}
\newcommand{\pkg}[1]{\textbf{#1}\xspace}
\newcommand{\code}[1]{\texttt{#1}\xspace}
\usepackage{booktabs}
\usepackage{hyperref}
\usepackage{url}

\usepackage[caption = false]{subfig}

\begin{document}

\title{\pkg{distr6}: R6 Object-Oriented Probability Distributions Interface in R}
\author{Raphael Sonabend and Franz J. Kir\'aly}

\maketitle

\abstract{
\pkg{distr6} is an object-oriented (OO) probability distributions interface leveraging the extensibility and scalability of R6, and the speed and efficiency of \pkg{Rcpp}. Over 50 probability distributions are currently implemented in the package with `core' methods including density, distribution, and generating functions, and more `exotic' ones including hazards and distribution function anti-derivatives. In addition to simple distributions, \pkg{distr6} supports compositions such as truncation, mixtures, and product distributions. This paper presents the core functionality of the package and demonstrates examples for key use-cases. In addition this paper provides a critical review of the object-oriented programming paradigms in R and describes some novel implementations for design patterns and core object-oriented features introduced by the package for supporting \pkg{distr6} components.
}


\section{Introduction}
\label{sec:tools_distr6_intro}
Probability distributions are an essential part of data science, underpinning models, simulations, and inference. Hence, they are central to computational data science. With the advent of modern machine learning and AI, it has become increasingly common to adopt a conceptual model where distributions are considered objects in their own right, as opposed to primarily represented through distribution defining functions (e.g., cdf, pdf), or random samples.

An important distinction to keep in mind is between random variables (that can be sampled from) and probability distributions. \pkg{distr6} is an interface for probability distributions and supports construction, manipulation, composition, and querying of parametrised simple and composite distributions. \pkg{distr6} is not an interface for random variables and therefore procedures such as sampling and inference are out of scope.

\subsection{The conceptual model: probability distributions and random variables}

We continue by explaining our conceptual model of probability distributions underpinning the design of \pkg{distr6}, and delineate it from common conceptualization of random variables. A full mathematical definition of the conceptual model is given in the next section; this section contains an intuitive introduction.

First, we invite the reader to recall some common mathematical objects and recognize that these are related but conceptually distinct:
\begin{itemize}
\itemsep-0.2em
\item a random variable, distributed according to a certain distribution, e.g., $X\sim \mbox{Normal}(0,1)$
\item the cdf of that random variable $X$, usually denoted by $F_X$, a function $F_X:\mathbb{R}\rightarrow [0,1]$
\item the pdf of that random variable $X$, often denoted by $f_X$, a function $f_X:\mathbb{R}\rightarrow [0,\infty )$
\item the distribution according to which $X$ is distributed - often called `the law of' $X$. This can be represented by multiple mathematical objects, such as the cdf $F_X$ or the pdf $f_X$. We will call this distribution $d$. Note that $d$ is not identical to either these representation functions.
\end{itemize}

\begin{figure}[ht!]
\centering
\subfloat[Discrete Uniform Random Variable]{\includegraphics[width = 8cm]{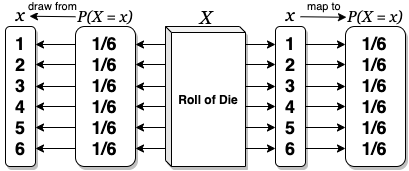}}
\\
\subfloat[Discrete Uniform Probability Distribution]{\includegraphics[width = 8cm]{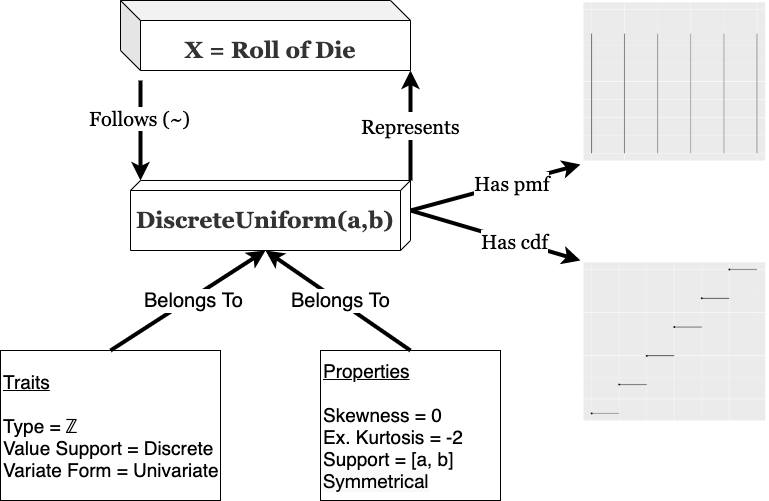}}
\caption[Discrete Uniform Random Variable and Distribution]{Schematic representation of concepts ``probability distribution'' and ``random variable''. (a) Example of a random variable following a Discrete Uniform distribution. A random variable is an object in its own right, modelling the process of a random experiment. It \emph{has} a probability distribution describing the nature of the experiment (in the case of uniform: the masses on the 6 outcomes), and can be sampled from, to obtain realizations. (b) Example of a probability distribution, the Discrete Uniform distribution. A probability distribution is an object in its own right and can serve as a template for random experiments, represented as random variables. Properties and traits such as value support, mean, variance, and functions such as pmf, cdf, are properties and traits of the probability distribution, not (direct) properties of a random variables, even if random variables can follow a given distribution.}
\label{fig:distr6_discreteuniform}
\end{figure}

Critically, we highlight that random variables and distributions are neither identical objects nor concepts. A random variable $X$ \emph{has} distribution $d$, and multiple random variables may be distributed according to $d$. Further, random variables are sampled from, while the distribution is only a description of probabilities for $X$. Thus, $X$ and $d$ are not identical objects. Figure~\ref{fig:distr6_discreteuniform} visually summarizes these differences.

As a possible logical consequence of the above, we adopt the conceptual model that a distribution is an abstract object, which:
\begin{itemize}
\itemsep-0.2em
\item Has multiple defining representations, for example through cdf and possibly through pdf, but is not identical with any of these representations
\item Possesses traits, such as being absolutely continuous over the Reals, and properties, such as skewness and symmetry.
\item Can be used to define sampling laws of random variables, but is not conceptually identical with a random variable.
\end{itemize}

Abstracting distributions as objects from multiple, non-identical, representations (random variables), introduces major consequences for the conceptual model:

\begin{enumerate}
\itemsep-0.2em
\item[(i)] It lends itself naturally to a class-object representation, in the computer scientific sense of object oriented programming. Abstract distributions become classes, concrete distributions are objects, and distribution defining functions are methods of these classes. Random variables are a separate type of object.
\item[(ii)] It strongly suggests adoption of mathematical conceptualization and notation which cleanly separates distributions from random variables and distribution defining functions - in contrast to common convention where random variables or random sampling takes conceptual primacy above all.
\item[(iii)] It allows clean formulation of algorithmic manipulations involving distributions, especially higher-order constructs (truncation, huberization, etc.), as well as clean mathematical definitions.
\end{enumerate}

\subsection{Distributions as software objects and mathematical objects}

In \pkg{distr6}, distributions are first-class objects subject to an object oriented class-object representation. For example, a discrete uniform distribution (fig.~\ref{fig:distr6_discreteuniform}b) is a `class' with traits such as type (Naturals), and variate form (univariate). With a given parametrization, this becomes an `object' with properties including symmetry and support. An alternative definition to the conceptual model of distributions is now provided.
\\\\
On the mathematical level, we again consider distributions as objects in their own right, not being identical with a cdf, pdf, or measure, but instead `having' these as properties.

For a set $\mathcal{Y}$ (endowed with suitable topology), we define Distr$(\mathcal{Y})$ as a set containing formal objects $d$ which are in bijection to (but not identical with) probability measures over $\mathcal{Y}$. Elements of Distr$(\mathcal{Y})$ are called distributions over $\mathcal{Y}$.
We further define formal symbols which, in case of existence, denote `aspects' that such elements have, in the following way: the symbol $d.F$, for example, denotes the cdf of $d$, which is to be read as the `$F$' of $d$, with $F$ in this case to be read as a modifier to a standard symbol $d$, rather than a fixed, bound, or variable symbol. In this way, we can define:
\begin{enumerate}
\itemsep-0.2em
\item[(i)] $d.F$ for the cdf of $d$. This typically exists if $\mathcal{Y}\subseteq \mathbb{R}^n$ for some $n$, in which case $d.F$ is a function of type $d.F: \mathbb{R}^n \rightarrow [0,1]$.
\item[(ii)] $d.f$ for the pdf of $d$. This exists if $\mathcal{Y}\subseteq \mathbb{R}^n$, and the distribution $d$ is absolutely continuous over $\mathcal{Y}$. In this case, $d.f$ is a function of type $d.f: \mathbb{R}^n \rightarrow [0,\infty)$.
\item[(iii)] $d.P$ for the probability measure that is in bijection with $d$. This is a function $d.P: \mathcal{F} \rightarrow [0,1]$ where $\mathcal{F}$ is the set of measurable sub-sets of $\mathcal{Y}$.
\end{enumerate}

We would like to point out that the above is indeed a full formal mathematical definition of our notion of distribution. While distributions, defined this way, are not identical with any of the conventional mathematical objects that define them (cdf, pdf, measures), they are conceptually, formally, and notationally well-defined. Similarly, the aspects ($d.F$, $d.f$, etc) are also well-defined, since they refer to one of the conventional mathematical objects which are well-specified in dependence of the distribution (in case of existence).

This notation provides a more natural and clearer separation of distribution and random variables and allows us to talk about and denote concepts such as `the cdf of any random variable following the distribution $d$' with ease ($d.F$), unlike classical notation that would see one define $X\sim d$ and then write $F_X$. Our notation more clearly follows the software implementation of distributions.

For example, in \pkg{distr6}, the code counterpart to defining a distribution $d$ which is Gaussian with mean $1$ and variance $2$ is
\begin{verbatim}
> d <- Normal$new(1, 2)
\end{verbatim}
The pdf and cdf of this Gaussian distribution evaluated at $2$ are obtained in code as
\begin{verbatim}
> d$pdf(2)
> d$cdf(2)
\end{verbatim}
which evaluates to `numerics' that represent the real numbers $d.f(2)$ and $d.F(2)$.
\\\\
The consideration of distributions as objects, and their conceptual distinction from random variables as objects, notably differs from conceptualization in R \pkg{stats}, which implements both distribution and random variable methods by the `\code{dpqr}' functions. Whilst this may allow very fast generation of probabilities and values, there is no support for querying and inspection of distributions as objects. By instead treating the \code{dpqr} functions as methods that belong to a distribution object, \pkg{distr6} encapsulates all the information in R \pkg{stats} as well as distribution properties, traits, and other important mathematical methods. The object orientation principle that defines the architecture of \pkg{distr6} is further discussed throughout this manuscript.

Treating distributions as objects is not unique to this package. Possibly the first instance of the object oriented conceptualization is the \pkg{distr} \citep{pkgdistr} family of packages. \pkg{distr6} was designed alongside the authors of \pkg{distr} in order to port some of their functionality from S4 to R6.

 \pkg{distr6} is the first such package to use the `class' object-oriented paradigm \pkg{R6} \citep{pkgR6}, with other distribution related packages using S3 or S4. The choice of R6 over S3 and S4 is discussed in detail in section \ref{sec:tools_distr6_oop}. This choice allows \pkg{distr6} to fully leverage the conceptual model, and make use of core R6 functionality. As well as introducing fundamental object-oriented programming (OOP) principles such as abstract classes, and tried and tested design patterns \citep{Gamma1996} including decorators, wrappers, and compositors (see section \ref{sec:tools_distr6_oop_cont}).
\\\\
Besides an overview to \pkg{distr6}'s novel approach to probability distributions in R, this paper also presents a formal comparison of the different OOP paradigms, while detailing the use of design patterns relevant to the package.

\subsection{Motivating example: Higher-order distribution constructs}

The strength of the object oriented approach, both on the algorithmic and mathematical side, lies in its ability to efficiently express higher-order constructs and operations: actions between distributions, resulting in new distributions. One such example is mixture distributions (also known as spliced distributions). In the \pkg{distr6} software interface, a \code{MixtureDistribution} is a higher-order distribution depending on two or more other distributions. For example take a uniform mixture of two distributions \texttt{distr1} and \texttt{distr2}:

\begin{verbatim}
> my_mixt <- MixtureDistribution$new(list(distr1, distr2))
\end{verbatim}
Internally, the dependency of the constructs on the components is remembered so that \texttt{my\_mixt} is not only evaluable for \texttt{cdf} (and other methods), but also carries a symbolic representation of its construction and definition history in terms of \texttt{distr1} and \texttt{distr2}.
\\\\
On the mathematical side, the object oriented formalism allows clean definitions of otherwise more obscure concepts, for example the mixture distribution is now defined by:

For distributions $d_1,\dots,d_m$ over $\mathbb{R}^n$ and weights $w_1,\dots, w_m$, we define the mixture of $d_1,\dots, d_m$ with weights $w_1,\dots, w_m$ to be the unique distribution $\tilde{d}$ such that $\tilde{d}.F(x) = \sum_{i=1}^m w_i\cdot d_i.F(x)$ for any $x\in \mathbb{R}^n$. Note the added clarity by defining the mixture on the distribution $d_i$, i.e., a first-order concept in terms of distributions.

\section{Related software}
\label{sec:tools_distr6_rel}

This section provides a review to other related software that implement probability distributions, this is focused on, but not limited to, software in R.

\paragraph{R stats, actuar, and extraDistr}
The core R programming language consists of packages for basic coding and maths as well as the \pkg{stats} package for statistical functions. \pkg{stats} contains 17 common probability distributions and four lesser-known distributions. Each distribution consists of (at most) four functions: \code{dX, pX, qX, rX} where \code{X} represents the distribution name. These correspond to the probability density/mass, cumulative distribution, quantile (inverse cumulative distribution) and simulation functions respectively. Each is implemented as a separate function, written in C, with both inputs and outputs as numerics. The strength of these functions lies in their speed and efficiency, there is no quicker way to find, say, the pdf of a Normal distribution than to run the \code{dnorm} function from \pkg{stats}. However, this is the limit of the package in terms of probability distributions. As there is no designated distribution object, there is no way to query results from the distributions outside of the `\code{dpqr}' functions.
\\\\
Several R packages implement \code{dpqr} functions for extra probability distributions. Of particular note are the \pkg{extraDistr} \citep{pkgextradistr} and \pkg{actuar} \citep{pkgactuar} packages that add over 60 distributions between them. Both of these packages are limited to \code{dpqr} functions and therefore have the same limits as R \pkg{stats}.

\paragraph{distr}
The \pkg{distr} package was the first package in R to implement an object-oriented interface for distributions, using the S4 object-oriented paradigm. \pkg{distr} tackles the two fundamental problems of \pkg{stats} by introducing distributions as objects that can be stored and queried. These objects include important statistical results, for example the expectation, variance and moment generating functions of a distribution. The \pkg{distr} family of packages includes a total of five packages for object-oriented distributions in R. \pkg{distr} has two weaknesses that were caused by using the S4 paradigm. Firstly, the package relies on inheritance, which means that large class trees exist for every object and extensibility is therefore non-trivial. The second weakness is that S4 objects are not referred to by `pointers' but instead copies. This means that a simple mixture of two distributions is just under 0.5Mb in size (relatively quite large).

\paragraph{distributions3 and distributional}
The \pkg{distributions3} package \citep{pkgdistributions3} defines distributions as objects using the S3 paradigm. However, whilst \pkg{distributions3} treats probability distributions as S3 objects, it does not add any properties, traits, or methods and instead uses the objects solely for \code{dpqr} dispatch. In comparison to \pkg{distr}, the \pkg{distributions3} package provides less features for inspection or composition. More recently, \pkg{distributional} \citep{pkgdistributional} builds on the \pkg{distributions3} framework (common authors exist between the two) to focus on the vectorization of probability distributions coded as S3 objects. Similarly to \pkg{distr6}, the primary use-case of this package is for predictive modelling of distributions as objects.

\paragraph{mistr}
The \pkg{mistr} package \citep{pkgmistr} is another recent distributions package, which is also influenced by \pkg{distr}. The sole focus of \pkg{mistr} is to add a comprehensive and flexible framework for composite models and mixed distributions. Similarly to the previous two packages, \pkg{mistr} implements distributions as S3 objects.

\paragraph{Distributions.jl}
Despite not being a package written in R, the Julia \pkg{Distributions.jl} \citep{pkgdistributions} package provided inspiration for \pkg{distr6}. \pkg{Distributions.jl} implements distributions as objects with statistical properties including expectation, variance, moment generating and characteristic functions, and many more. This package uses multiple inheritance for `valueSupport' (discrete/continuous) and `variateForm' (univariate/multivariate/matrixvariate). Every distribution inherits from both of these, e.g. a distribution can be `discrete-univariate', `continuous-multivariate', `continuous-matrixvariate', etc. The package provides a unified and user-friendly interface, which was a helpful starting point for \pkg{distr6}.


\section{Design principles}
\label{sec:tools_distr6_principles}

\pkg{distr6} was designed and built around the following principles.

\begin{itemize}
\item[D1)] \textbf{Unified interface} The package is designed such that all distributions, no matter how complex, have an identical user-facing interface. This helps make the package easy to navigate and the documentation simple to read. Moreover it minimises any confusion resulting from using multiple distributions. A clear inheritance structure also allows wrappers and decorators to have the same methods as distributions, which means even complex composite distributions should be intuitive to use. Whether a user constructs a simple Uniform distribution, or a mixture of 100 Normal distributions, the same methods and fields are seen in both objects.


\item[D2)] \textbf{Separation of core/exotic and numerical/analytic} Via abstraction and encapsulation, core statistical results (common methods such as mean and variance) are separated from `exotic' ones (less common methods such as anti-derivatives and p-norms). Similarly, implemented distributions only contain analytic results; users can impute numerical results using decorators. This separation has several benefits including: 1) for predictive modelling with/of distributions, numerical results can take longer to compute than analytical results and the difference between precision of analytical and numerical results can be substantial in the context of automated modelling, separation allows these differences to be highlighted and controlled; 2) separating numerical results allows an expanded interface for users to fine tune and set their own parameters for how numerical results are computed; 3) a less-technical user can guarantee precision of results as they are unlikely to use numerical decorators; 4) a user has access to the most important distribution methods immediately after construction but is not overwhelmed by many `exotic' methods that they may never use. Use of decorators and wrappers allow the user to manually expand the interface at any time. For example a user can choose between an undecorated Binomial distribution, with common methods such as mean and variance, or they can decorate the distribution to additionally gain access to survival and hazard functions.

\item[D3)] \textbf{Inheritance without over-inheritance} The class structure stems from a series of a few abstract classes with concrete child classes, which allows for a sensible, but not over-complicated, inheritance structure. For example all implemented distributions inherit from a single parent class so common methods can be unified and only coded once; note there is no separation of distributions into `continuous' and `discrete' classes. By allowing extension of classes by decorators and wrappers, and not solely inheritance, the interface is highly scalable and extensible. By `scalability' we refer to the interface's ability to grow to a large scale without additional overheads, the decorator and wrapper patterns on top of the R6 paradigm allow a (theoretically) unlimited number of distributions, wrappers, and methods without computational difficulty. By `extensibility' we refer to the ability to extend the interface, again this is made possible by clean abstraction of distributions, wrappers, core methods, and extra methods in decorators. All decorators and wrappers in \pkg{distr6} stem from abstract classes, which in turn inherit from the \code{Distribution} super-class. In doing so, any method of expanding an object's interface in \pkg{distr6} (i.e. via decorators, wrappers or inheritance) will automatically lead to an interface that inherits from the top-level class, maintaining the principle of a unified interface (D1).

\item[D4)] \textbf{Inspection and manipulation of multiple parameterisations} The design process identified that use of distributions in R \pkg{stats} is inflexible in that in the majority of cases, only one parameterisation of each distribution is allowed. This can lead to isolating users who may be very familiar with one parameterisation but completely unaware of another. For example the use of the \textit{precision} parameter in the Normal distribution is typically more common in Bayesian statistics whereas using the \textit{variance} or \textit{standard deviation} parameters are more common in frequentist statistics. \pkg{distr6} allows the user to choose from multiple parameterisations for all distributions (where more than one parameterisation is possible/known). Furthermore, querying and updating of any parameter in the distribution is allowed, even if it was not specified in construction (section \ref{sec:tools_distr6_api}). This allows for a flexible parameter interface that can be fully queried and modified at any time.

\item[D5)] \textbf{Flexible interfacing for technical and non-technical users} Throughout the design process, it was required that \pkg{distr6} be accessible to all R users. This was a challenge as R6 is a very different paradigm from S3 and S4. To reduce the learning curve, the interface is designed to be as  user-friendly and flexible as possible. This includes: 1) a `sensible default principle' such that all distributions have justified default values; 2) an `inspection principle' with functions to list all distributions, wrappers, and decorators. As discussed in (D2), abstraction and encapsulation allow technical users to expand any distribution's interface to be as arbitrarily complex as they like, whilst maintaining a minimal representation by default. Where possible defaults are `standard' distributions, i.e. with location $0$ and scale $1$, otherwise sensible defaults are identified as realistic scenarios, for example \code{Binomial(n = 10, p = 0.5)}.

\item[D6)] \textbf{Flexible OO paradigms} Following from (D5), we identified that R6 is still relatively new in R with only $314$ out of $16, 050$ packages depending on it (as of July 2020). Therefore this was acknowledged and taken into account when building the package. R6 is also the first paradigm in R with the dollar-sign notation (though S4 uses `@' notation) and with a proper construction method. Whilst new users are advised to learn the basics of R6, S3 compatibility is available for all common methods via \pkg{R62S3} \citep{pkgr62s3}. Users can therefore decide on calling a method via dollar-sign notation or dispatch, the example below demonstrates `piping' and S3. As the core package is built on R6, the thin-wrappers provided by \pkg{R62S3} do not compromise the above design principles.
\end{itemize}

\begin{verbatim}
> library(magrittr)
> N <- Normal$new(mean = 2)
> N %>%
+   setParameterValue(mean = 1) %>%
+   getParameterValue("mean")
[1] 1
> pdf(N, 1:4)
[1] 0.398942280 0.241970725 0.053990967 0.004431848
\end{verbatim}

\section{Overview to functionality and API}
\label{sec:tools_distr6_api}

\pkg{distr6} 1.4.3 implements 56 probability distributions, including 11 probability kernels. Individual distributions are modelled via classes that inherit from a common interface, implemented in the abstract \code{Distribution} parent class. The \code{Distribution} class specifies the abstract distribution interface for parameter access, properties, traits, and methods, such as a distribution's pdf or cdf. The most important interface points are described in Section~\ref{sec:tools_distr6_api_distr}

\begin{figure}[h]
\centering
\begin{tikzpicture}
\umlclass{Distribution}{+ name : string \\+ short\_name : string \\ + description : string \\ - .properties : list \\ - \underline{.traits} : list}{+ public\_accessors() \\ + \code{dpqr} methods() \\ + generic\_math\_stat\_methods() \\ + ParameterSet\_getters\_setters() \\ + validation\_methods()}
\end{tikzpicture}
\end{figure}

Concrete distributions, kernels, and wrappers are the grandchildren of \code{Distribution}, and children of one of the mid-layer abstract classes:
\begin{itemize}
\itemsep-0.2em
\item \code{SDistribution}, which models abstract, generic distributions. Concrete distributions, such as \code{Normal} which models the normal distribution, inherit from \code{SDistribution}.
\item \code{Kernel}, which models probability kernels, such as \code{Triangular} and \code{Epanechnikov}. Probability kernels are absolutely continuous distributions over the Reals, with assumed mean 0 and variance 1.
\item \code{DistributionWrapper}, which is an abstract parent for higher-order operations on distributions, including compositions, that is, operations that create distributions from other distributions, such as truncation or mixture.
\item \code{DistributionDecorator}, whose purpose is supplementing methods to distributions in the form of a decorator design pattern, this includes methods such as integrated cdf or squared integrals of distribution defining functions.
\end{itemize}

\begin{figure}[h]
\centering
\begin{tikzpicture}
\begin{umlpackage}{distr6}
\umlsimpleclass{Distribution}{}{}
\umlsimpleclass[x = 0.5, y = -2, type = abstract]{SDistribution}{}{}
\umlsimpleclass[x = -2.5, y = -2, type = abstract]{Kernel}{}{}
\umlsimpleclass[x = -6, y = -2, type = abstract]{DistributionDecorator}{}{}
\umlsimpleclass[x = 4.5, y = -2, type = abstract]{DistributionWrapper}{}{}

\umlsimpleclass[x = 0.5, y = -4]{SDistributionX}{}{}
\umlsimpleclass[x = -2.5, y = -4]{KernelX}{}{}
\umlsimpleclass[x = -6, y = -4]{DecoratorX}{}{}
\umlsimpleclass[x = 4.5, y = -4]{WrapperX}{}{}

\umlinherit[geometry=-|]{DistributionDecorator}{Distribution}
\umlinherit[geometry=-|]{SDistribution}{Distribution}
\umlinherit[geometry=-|]{Kernel}{Distribution}
\umlinherit[geometry=-|]{DistributionWrapper}{Distribution}

\umlinherit[geometry=-|]{DecoratorX}{DistributionDecorator}
\umlinherit[geometry=-|]{WrapperX}{DistributionWrapper}
\umlinherit[geometry=-|]{KernelX}{Kernel}
\umlinherit[geometry=-|]{SDistributionX}{SDistribution}

\umluniaggreg[geometry=|-, anchors=90 and 180]{DistributionDecorator}{Distribution}
\umluniaggreg[geometry=|-, anchors=90 and 0]{DistributionWrapper}{Distribution}

\end{umlpackage}
\end{tikzpicture}
\end{figure}

The UML diagram above visualises the key class structure of \pkg{distr6} including the concrete \code{Distribution} parent class, from which all other classes in the package inherit from (with the exception of the \code{ParameterSet}). These abstract classes allow simple extensibility for concrete sub-classes.

\subsection{The Distribution interface}
\label{sec:tools_distr6_api_distr}
The base, or top-level, class in \pkg{distr6} is the \code{Distribution} class. Its primary function is to act as a parent class for the implemented probability distributions and higher-order compositions, it is also utilised for creation of custom distributions. By design, any distribution already implemented in \pkg{distr6} will have the same interface as a user-specified custom distribution, ensuring (D1) is upheld. A table of the most important methods for a distribution are in Table \ref{tab:methods} alongside their meaning and definitions (mathematical if possible). The two use-cases for the \code{Distribution} class are discussed separately.

\begin{table}
\centering
\begin{tabular}{p{4cm}p{9cm}}
\toprule
\textbf{Method} & \textbf{Description/Definition} \\
\midrule
\code{pdf/cdf/quantile/rand} & \code{dpqr} functions. \\
\code{mean} & $d.\mu = \mathbb{E}[X]$ \\
\code{variance}& $d.\sigma^2 = \mathbb{E}[(X - d.\mu)^2]$ \\
\hline
\code{traits} & List including value support (discrete/continuous/mixed); variate form (uni-/multi-/matrixvariate); type (mathematical domain). \\
\code{properties} & List including skewness ($\mathbb{E}((X - d.\mu)/d.\sigma)^3)$) and symmetry (boolean). \\
\hline
\code{get/setParameterValue} & Getters and setters for parameter values. \\
\code{parameters} & Returns the internal parameterisation set. \\
\hline
\code{print/summary} & Representation functions, summary includes distribution properties and traits. \\
\bottomrule
\end{tabular}
\caption{Common methods available to all classes inheriting from \code{Distribution}. Columns are method names and either mathematical definition or description of method. For mean and variance, $d.\mu$, $d.\sigma$, and $d.\sigma^2$ represent the methods mean, standard deviation, and variance associated with distribution $d$. The symbol $X$ denotes a random variable with distribution $d$. Horizontal lines separate mathematical, property, parameter, and representation methods.}\label{tab:methods}
\end{table}

\paragraph{Distribution for inheritance}
It is anticipated that the majority of \pkg{distr6} users will be using the package for the implemented distributions and kernels. With this in mind, the \code{Distribution} class defines all variables and methods common to all child classes. The most important of these are the common analytical expressions and the \code{dpqr} public methods. Every concrete implemented distribution/kernel has identical public \code{dpqr} methods that internally call private \code{dpqr} methods. This accounts for inconsistencies occurring from packages returning functions in different formats and handling errors differently; a problem most prominent in multivariate distributions. Another example is handling of non-integer values for discrete distributions, in some packages this returns $0$, in others the value is rounded down, and in others an error is returned. The \code{dpqr} functions for all distributions have unified validation checks and return types (\code{numeric} or \code{data.table}). In line with base R and other distribution packages, \pkg{distr6} implements a single \code{pdf} function to cover both probability mass and probability density functions.


\begin{verbatim}
> Normal$new()$pdf(1:2)
[1] 0.24197072 0.05399097
> Binomial$new()$cdf(1:2, lower.tail = FALSE, log.p = TRUE, simplify = FALSE)
         Binom
1: -0.01080030
2: -0.05623972
\end{verbatim}

A key design principle in the package is separation of analytical and numerical results (D2), which is ensured by only including analytical results in implemented distributions. Missing methods in a distribution therefore signify that no closed-form expression for the method is available, however all can be numerically estimated with the \code{CoreStatistics} decorator (see section \ref{sec:tools_distr6_api_decor}). Ideally, all distributions will include analytical methods for the following: probability density/mass function (\code{pdf}),  cumulative distribution function (\code{cdf}), inverse cumulative distribution function/quantile function (\code{quantile}), simulation function (\code{rand}), mean, variance, skewness, (excess) kurtosis, and entropy of the distribution (\code{mean, variance, skewness, kurtosis, entropy}), as well as the moment generating function (\code{mgf}), characteristic function (\code{cf}), and probability generating function (\code{pgf}). Speed is currently a limitation in \pkg{distr6} but the use of \pkg{Rcpp} \citep{pkgrcpp} in all \code{dpqr} functions helps mitigate against this.

The fourth design principle of \pkg{distr6} ensures that multiple parameterisations of a given distribution can be both provided and inspected at all times. For example the Normal distribution can be parametrised in terms of variance, standard deviation, or precision. Any of which can be called in construction with other parameters updated accordingly. If conflicting parametrisations are provided then an error is returned. By example,

\begin{verbatim}
# set precision, others updated automatically
> Normal$new(prec = 4)
Norm(mean = 0, var = 0.25, sd = 0.5, prec = 4)
# try and set both precision and variance, results in error
> Normal$new(var = 1, prec = 2)
Error in FUN(X[[i]], ...) :
  Conflicting parametrisations detected. Only one of {var, sd, prec} should be given.
\end{verbatim}

The same principle is used for parameter setting. The methods \code{getParameterValue} and \\ \code{setParameterValue} are utilised for getting and setting parameter values respectively. The former takes a single argument, the parameter name, and the second a named list of arguments corresponding to the parameter name and the value to set. The example below demonstrates this for a Gamma distribution. Here the distribution is constructed, the shape parameter is queried, both shape and rate parameters are updated and the latter queried, finally the scale parameter is set which auto-updates the rate parameter.
\begin{verbatim}
> G <- Gamma$new(shape = 1, rate = 1)
> G$getParameterValue("shape")
[1] 1
> G$setParameterValue(shape = 1, rate = 2)
> G$getParameterValue("rate")
[1] 2
> G$setParameterValue(scale = 2)
> G$getParameterValue("rate")
[1] 0.5
\end{verbatim}

Distribution and parameter domains and types are represented by mathematical sets, implemented in \pkg{set6} \citep{pkgset6}. This allows for clear representation of infinite sets and most importantly for internal containedness checks. For example all public \code{dpqr} methods first call the \code{\$contains} method in their respective \code{type} and return an error if any points are outside the distribution's domain. As \pkg{set6} uses \pkg{Rcpp} for this method, these come at minimal cost to speed.

\begin{verbatim}
> B <- Binomial$new()
> B$pdf(-1)
Error in B$pdf(-1) :
  Not all points in {-1} lie in the distribution domain (N0).
\end{verbatim}

These domains and types are returned along with other important properties and traits in a call to \code{properties} and \code{traits} respectively, this is demonstrated below for the Arcsine distribution.

\begin{verbatim}
> A <- Arcsine$new()
> A$properties
$support
[0,1]


$symmetry
[1] "symmetric"

> A$traits
$valueSupport
[1] "continuous"

$variateForm
[1] "univariate"

$type
R
\end{verbatim}

\paragraph{Extending \pkg{distr6} with custom distributions}
Users of \pkg{distr6} can create temporary custom distributions using the constructor of the \code{Distribution} class directly. Permanent extensions, e.g., as part of an R package, should create a new concrete distribution as a child of the \code{SDistribution} class.

The \code{Distribution} constructor is given by

\begin{verbatim}
Distribution$new(name = NULL, short_name = NULL, type = NULL, support = NULL,
+ symmetric = FALSE, pdf = NULL, cdf = NULL, quantile = NULL, rand = NULL,
+ parameters = NULL, decorators = NULL, valueSupport = NULL, variateForm = NULL,
+ description = NULL)
\end{verbatim}

The \code{name} and \code{short\_name} arguments are identification for the custom distribution used for printing. \code{type} is a trait corresponding to scientific type (e.g. Reals, Integers,...) and \code{support} is the property of the distribution support. Distribution parameters are passed as a \code{ParameterSet} object, this defines each parameter in the distribution including the parameter default value and support. The \code{pdf/cdf/quantile/rand} arguments define the corresponding methods and are passed to the private \code{.pdf/.cdf/.quantile/.rand}  methods, as above the public methods are already defined and ensure consistency in each function. At a minimum users have to supply the distribution \code{name}, \code{type} and either \code{pdf} or \code{cdf}, all other information can be numerically estimated with decorators (see section \ref{sec:tools_distr6_api_decor}).

\begin{verbatim}
> d <- Distribution$new(name = "Custom Distribution", type = Integers$new(),
+                       support = Set$new(1:10),
+                       pdf = function(x) rep(1/10, length(x)))
> d$pdf(1:3)
[1] 0.1 0.1 0.1
\end{verbatim}

\subsection{DistributionDecorator}
\label{sec:tools_distr6_api_decor}
Decorators add functionality to classes in object-oriented programming. These are not natively implemented in R6 and this novel implementation is therefore discussed further in section \ref{sec:tools_distr6_oop_cont}. Decorators in \pkg{distr6} are only `allowed' if they have have at least three methods and cover a clear use-case, this prevents too many decorators bloating the interface. However by their nature, they are lightweight classes that will only increase the methods in a distribution if explicitly requested by a user. Decorators can be applied to a distribution in one of three ways:
\\\\
In construction:
\begin{verbatim}
> N <- Normal$new(decorators = c("CoreStatistics", "ExoticStatistics"))
\end{verbatim}

\noindent Using the \code{decorate()} function:
\begin{verbatim}
> N <- Normal$new()
> decorate(N, c("CoreStatistics", "ExoticStatistics"))
\end{verbatim}

\noindent Using the \code{\$decorate} method inherited from the \code{DistributionDecorator} super-class:
\begin{verbatim}
> N <- Normal$new()
> ExoticStatistics$new()$decorate(N)
\end{verbatim}

The first option is the quickest if decorators are required immediately. The second is the most efficient once a distribution is already constructed. The third is the closest method to true OOP but does not allow adding multiple decorators simultaneously.
\\\\
Three decorators are currently implemented in \pkg{distr6}, these are briefly described.

\paragraph{CoreStatistics} This decorator imputes numerical functions for common statistical results that could be considered core to a distribution, e.g. the mean or variance. The decorator additionally adds generalised expectation (\code{genExp}) and moments (\code{kthmoment}) functions, which allow numerical results for functions of the form $\mathbb{E}[f(X)]$ and for crude/raw/central $K$ moments. The example below demonstrates how the \code{decorate} function exposes methods from the \code{CoreStatistics} decorator to the Normal distribution object.

\begin{verbatim}
> n <- Normal$new(mean = 2, var = 4)
> n$kthmoment(3, type = "raw")
Error: attempt to apply non-function
> decorate(n, CoreStatistics)
> n$kthmoment(3, type = "raw")
[1] 32
\end{verbatim}

\paragraph{ExoticStatistics} This decorator adds more `exotic' methods to distributions, i.e. those that are unlikely to be called by the majority of users. For example this includes methods for the p-norm of survival and cdf functions, as well as anti-derivatives for these functions. Where possible, analytic results are exploited. For example, this decorator can implement the survival function in one of two ways: either as i) $1$ minus the distribution cdf, if an analytic expression for the cdf is available; or ii) via numerical integration of the distribution.

\paragraph{FunctionImputation} This decorator imputes numerical expressions for the \code{dpqr} methods. This is most useful for custom distributions in which only the \code{pdf} or \code{cdf} is provided. Numerical imputation is implemented via \pkg{Rcpp}.

\subsection{Composite distributions}
\label{sec:tools_distr6_api_wrap}

Composite distributions - that is, distributions created from other distributions - are common in advanced usage. Examples for composites are truncation, mixture, or transformation of domain. In \pkg{distr6}, a number of such composites are supported. Implementation-wise, this uses the wrapper OOP pattern, which is not native to R6 but part of our extensions to R6, discussed in section \ref{sec:tools_distr6_oop_cont}.

As discussed above, wrapped distributions inherit from \code{Distribution} thus have an identical interface to any child of \code{SDistribution}, with the following minor differences:
\begin{itemize}
\itemsep-0.2em
\item The \code{wrappedModels} method provides a unified interface to access any component distribution.
\item Parameters are still accessed via the same method but stored in a \code{ParameterSetCollection} object instead of a \code{ParameterSet}, thus allowing efficient representation of composite and nested parameter sets.
\end{itemize}

Composition can be iterated and nested any number of times, consider the following example where a mixture distribution is created from two distributions that are in turn composites - a truncated Student T, and a huberized exponential - note too the parameter inspection and automatic prefixing of distribution `short names' to the parameters for identification:

%

\begin{verbatim}
> M <- MixtureDistribution$new(list(
+   truncate(StudentT$new(), lower = -1, upper = 1),
+   huberize(Exponential$new(), upper = 4)
+ ))
> M$parameters()


                id   value           support                      description
1:        mix_T_df       1                    R+           Degrees of Freedom
2: mix_trunc_lower      -1      R U {-Inf, +Inf}    Lower limit of truncation
3: mix_trunc_upper       1       R U {-Inf +Inf}    Upper limit of truncation
4:    mix_Exp_rate       1                    R+                 Arrival Rate
5:   mix_Exp_scale       1                    R+                        Scale
6:   mix_hub_lower       0      R U {-Inf, +Inf}  Lower limit of huberization
7:   mix_hub_upper       4      R U {-Inf, +Inf}  Upper limit of huberization
8:     mix_weights uniform     {uniform} U [0,1]              Mixture weights
\end{verbatim}

\paragraph{Implemented compositors}

We summarize some important implemented compositors (tables \ref{tab:comp1} and \ref{tab:comp2}), to illustrate the way composition is handled and implemented.

\begin{table}[h]
\begin{tabular}{cccc}
  \toprule
   \textbf{Class} & \textbf{Parameters} & \textbf{Type of $d$ } & \textbf{Components of $d$} \\\hline
  \code{TruncatedDistribution} & $a,b\in \mathbb{R}$ & $\mathbb{R}$ & $d'$, type $\mathbb{R}$ \\
  \code{HuberizedDistribution} & $a,b\in \mathbb{R}$ & $\mathbb{R},$ mixed & $d'$, type $\mathbb{R}$ \\
  \code{MixtureDistribution} & $w_i \in \mathbb{R}, \sum^n_{i=1} w_i = 1$ & $\mathbb{R}^n$ & $d'_i$, type $\mathbb{R}^n$ \\
  \code{ProductDistribution} & - & $\mathbb{R}^N, N = \sum_{i=1}^n n_i$ & $d'_i$ type $\mathbb{R}^{n_i}$ \\
  \bottomrule
\end{tabular}
\caption{\label{tab:comp1} Examples of common compositors implemented in \pkg{distr6} - parameters and type. Column 2 (parameters) lists the parameters that the composite has. Column 3 (type) states the type of the resultant distribution $d$ that is created when the class is constructed; this states the formal domain. Column 4 (components) states the number, names, and assumptions on the components, if any.}
\end{table}

\begin{table}[h]
\begin{tabular}{ccc}
  \toprule
 \textbf{Class} & $d.F(x)$ & $d.f(x)$ \\ \midrule
  \code{TruncatedDistribution} & $\frac{d'.F(x) - d'.F(a)}{d'.F(b) - d'.F(a)}$ & $\frac{d'.f(x)}{d'.P([a,b])}$ \\
  \code{HuberizedDistribution} & $d'.F(x) + \mathbbm{1}[x = b]\cdot d'.P(b)$ & (no pdf, since mixed) \\
  \code{MixtureDistribution} & $\sum_{i=1}^N w_i\cdot d'_i.F(x)$ & $\sum_{i=1}^N w_i\cdot d'_i.f(x)$ (if exists) \\
  \code{ProductDistribution} & $\prod_{i=1}^N d'_i.F(x)$ & $\prod_{i=1}^N d'_i.f(x)$ (if exists) \\
  \bottomrule
\end{tabular}
\caption{\label{tab:comp2} Common compositors implemented in \pkg{distr6} with mathematical definitions. Column 2 defines the resultant cdf, $d.F$, in terms of the component cdf, as implemented in the \code{cdf} method of the compositor in the same row. Column 3 defines the resultant pdf, $d.f$, in terms of the component pdf, as implemented in the \code{pdf} method of the compositor in the same row. Truncation (row 1) is currently implemented for the left-open interval $(a, b]$ only. For Huberization (row 2), the resultant distribution is in general not absolutely continuous and hence the pdf does not exist. $d'$ resp.~$d_i'$ are the component distributions, as defined in table \ref{tab:comp1}.}
\end{table}

Example code to obtain a truncated or huberized distribution is below. Here, we construct a truncated normal with truncation parameters -1 and 1, and a huberized Binomial with bounding parameters 2 and 5.

\begin{verbatim}
> TN <- truncate(Normal$new(), lower = -1, upper = 1)
> TN$cdf(-2:2)
[1] 0.0 0.0 0.5 1.0 1.0
> class(TN)
[1] "TruncatedDistribution" "DistributionWrapper" "Distribution" "R6"

> HB <- huberize(Binomial$new(), lower = 2, upper = 5)
> HB$cdf(1:6)
[1] 0.0000000 0.0546875 0.1718750 0.3769531 1.0000000 1.0000000
> HB$median()
[1] 5
\end{verbatim}

\paragraph{Vectorization of distributions}

A special feature of \pkg{distr6} is that it allows vectorization of distributions,  i.e. vectorised representation of multiple distributions in an array-like structure. This is primarily done for computational efficiency with general best R practice of vectorisation. Vectorisation of \pkg{distr6} distributions is implemented via the \code{VectorDistribution} which is logically treated as a compositor.

Mathematically, a \code{VectorDistribution} is simply a vector of component distributions $d_1,\dots, d_N$ that allows vectorised evaluation. Two kinds of vectorised evaluation are supported - paired and product-wise vectorization, which we illustrate in the case of cdfs.
\begin{itemize}
\itemsep-0.2em
\item Paired vectorised evaluation of the cdfs $d_1.F,\dots, d_N.F$ at numbers $x_1,\dots, x_N,$ yields a real vector $(d_1.F(x_1),\dots, d_N.F(x_N)$ via the \code{cdf} method.
\item Product vectorised evaluation of the cdfs $d_1.F,\dots, d_N.F$ at numbers $x_1,\dots, x_M,$ yields a real $(N\times M)$ matrix, with $(i,j)$-th entry $d_i.F(x_j)$.
\end{itemize}

In practical terms, paired evaluation is the evaluation of $N$ distributions at $N$ points (which may be unique or different). So by example for three distributions $d_1,d_2,d_3$, paired evaluation of their cdfs at $(x_1, x_2, x_3) = (4,5,6)$ respectively results in $(d_1.F(x_1), d_2.F(x_2), d_3.F(x_3)) = (d_1.F(4), d_2.F(5), d_3.F(6))$. In \pkg{distr6}:
\begin{verbatim}
> V <- VectorDistribution$new(distribution = "Normal", params = data.frame(mean = 1:3))
> V$cdf(4, 5, 6)
       Norm1     Norm2     Norm3
1: 0.9986501 0.9986501 0.9986501
\end{verbatim}

In contrast, product-wise evaluation evaluates $N$ distributions at the same $M$ points. Product-wise evaluation of the cdfs of $d_1,d_2,d_3$ at $(x_1, x_2, x_3) = (4,5,6)$ results in
\begin{equation*}
\begin{pmatrix}
d_1.F(x_1) & d_1.F(x_2) & d_1.F(x_3) \\
d_2.F(x_1) & d_2.F(x_2) & d_2.F(x_3) \\
d_3.F(x_1) & d_3.F(x_2) & d_3.F(x_3)
\end{pmatrix} =
\begin{pmatrix}
d_1.F(4) & d_1.F(5) & d_1.F(6) \\
d_2.F(4) & d_2.F(5) & d_2.F(6) \\
d_3.F(4) & d_3.F(5) & d_3.F(6)
\end{pmatrix}
\end{equation*}

In \pkg{distr6}:

\begin{verbatim}
> V <- VectorDistribution$new(distribution = "Normal", params = data.frame(mean = 1:3))
> V$cdf(4:6)
       Norm1     Norm2     Norm3
1: 0.9986501 0.9772499 0.8413447
2: 0.9999683 0.9986501 0.9772499
3: 0.9999997 0.9999683 0.9986501
\end{verbatim}

The \code{VectorDistribution} wrapper allows for efficient vectorisation across \textit{both} the distributions \textit{and} points to evaluate, which we believe is a feature unique to \pkg{distr6} among distribution frameworks in R. By combing product and paired modes, users can evaluate any distribution in the vector at any point. In the following example, Normal(1, 1) is evaluated at (1,2) and Normal(2, 1) is evaluated at (3,4):

\begin{verbatim}
> V <- VectorDistribution$new(distribution = "Normal", params = data.frame(mean = 1:2))
> V$pdf(1:2, 3:4)
       Norm1      Norm2
1: 0.3989423 0.24197072
2: 0.2419707 0.05399097
\end{verbatim}

Further, common composites such as \code{ProductDistribution} and \code{MixtureDistribution} inherit from \code{VectorDistribution}, allowing for efficient vector dispatch of pdf and cdf methods. Inheriting from \code{VectorDistribution} results in identical constructor and methods. Thus a minor caveat is that users could evaluate a product or mixture at different points for each distribution, which is not a usual use-case in practice.

Two different choices of constructors are provided, the first `\code{distlist}' constructor passes distribution \textit{objects} into the constructor, whereas the second passes a reference to the distribution \textit{class} along with the parameterisations. Therefore the first allows different types of distributions but is vastly slowly as the various methods have to be calculated individually, whereas the second only allows a single class of distribution at a time, but is much quicker in evaluation. In the example below, the mixture uses the second constructor and the product uses the first.

\begin{verbatim}
> M <- MixtureDistribution$new(distribution = "Degenerate",
+                              params = data.frame(mean = 1:10))
> M$cdf(1:5)
[1] 0.1 0.2 0.3 0.4 0.5
> class(M)
[1] "MixtureDistribution" "VectorDistribution"  "DistributionWrapper" "Distribution"
[5] "R6"

> P <- ProductDistribution$new(list(Normal$new(), Exponential$new(), Gamma$new()))
> P$cdf(1:5)
[1] 0.3361815 0.7306360 0.9016858 0.9636737 0.9865692
\end{verbatim}

\section{Design patterns and object-oriented programming}
\label{sec:tools_distr6_patterns}

This paper has so far discussed the API and functionality in \pkg{distr6}. This section discusses object-oriented programming (OOP), firstly a brief introduction to OOP and OOP in R and then the package's contributions to the field.

\subsection{S3, S4, and R6}
\label{sec:tools_distr6_oop}

R has four major paradigms for object-oriented programming: S3, S4, reference classes (R5), and most recently, R6. S3 and S4 are known as functional object-oriented programming (FOOP) paradigms whereas R5 and R6 move towards class object-oriented programming (COOP) paradigms (R6) \citep{Chambers2014}. One of the main differences (from a user-perspective) is that methods in COOP are associated with a class whereas in FOOP, methods are associated with generic functions. In the first case methods are called by first specifying the object and in the second, a dispatch registry is utilised to find the correct method to associate with a given object.

S3 introduces objects as named structures, which in other languages are often referred to as `typed lists'. These can hold objects of any type and can include meta-information about the object itself. S3 is the dominant paradigm in R for its flexibility, speed, and efficiency. As such it is embedded deep in the infrastructure of R and single dispatch is behind a vast majority of the base functionality, which is a key part of making R easily readable. S3 is a FOOP paradigm in which functions are part of a dispatch system and consist of a generic function that is external to any object and a specific method registered to a `class'. However the term `class' is slightly misleading as no formal class structure exists (and by consequence no formal construction or inheritance) and as such S3 is not a formal OOP language\footnote{http://adv-r.had.co.nz/OO-essentials.html}.

S4 formalises S3 by introducing: class-object separation, a clear notion of construction, and multiple inheritance \citep{Chambers2014}. S4 has more syntax for the user to learn and a few more steps in class and method definitions, as a result S4 syntax is not overly user-friendly and S3 is used vastly more than S4 \citep{Chambers2014}.

There is a big jump from S3 and S4 to R6 as they transition from functional- to class-object-oriented programming. This means new notation, semantics, syntax, and conventions. The key changes are: 1) introducing methods and fields that are associated with classes not functions; 2) mutable objects with copy-on-modify semantics; and 3) new dollar-sign notation. In the first case this means that when a class is defined, all the methods are defined as existing within the class, and these can be accessed at any time after construction. Methods are further split into \textit{public} and \textit{private}, as well as \textit{active bindings}; which incorporates the abstraction part of OOP. The mutability of objects and change to copy-on-modify means that to create an independent copy of an object, the new method \code{clone(deep = TRUE)} has to be used, which would be familiar to users who know more classical OOP but very different to most R users. Finally methods are accessed via the dollar-sign, and not by calling a function on an object.

Below the three paradigms are contrasted semantically with a toy example to create a `duck' class with a method `quack':

\begin{center}
\textbf{S3}
\end{center}
\begin{verbatim}
> quack <- function(x) UseMethod("quack", x)
> duck <- function(name) return(structure(list(name = name), class = "duck"))
> quack.duck <- function(x) cat(x$name, "QUACK!")
> quack(duck("Arthur"))
Arthur QUACK!
\end{verbatim}
\newpage
\begin{center}
\textbf{S4}
\end{center}
\begin{verbatim}
> setClass("duck", slots = c(name = "character"))
> setGeneric("quack", function(x) {
+  standardGeneric("quack")
+  })
> setGeneric("duck", function(name) {
+   standardGeneric("duck")
+  })
> setMethod("duck", signature(name = "character"),
+             definition = function(name){
+              new("duck", name = name)
+              })
> setMethod("quack",
+             definition = function(x) {
+             cat(x@name, "QUACK!")
+             })
> quack(duck("Ford"))
Ford QUACK!
\end{verbatim}
\begin{center}
\textbf{R6}
\end{center}
\begin{verbatim}
> duck <- R6::R6Class("duck", public = list(
+   initialize = function(name) private$.name = name,
+   quack = function() cat(private$.name, "QUACK!")),
+   private = list(.name = character(0)))
> duck$new("Zaphod")$quack()
Zaphod QUACK!
\end{verbatim}

The example clearly highlights the extra code introduced by S4 and the difference between the S3 dispatch and R6 method system.

\paragraph{Comparing the paradigms}
There is no doubt that R6 is the furthest paradigm from conventional R usage and as such there is a steep learning curve for the majority of R users. However R6 will be most natural for users coming to R from more traditional OOP languages. In contrast, S3 is a natural FOOP paradigm that will be familiar to all R users (even if they are not aware that S3 is being used). S4 is an unfortunate midpoint between the two, which whilst being very useful, is not particularly user-friendly in terms of programming classes and objects.

\pkg{distr} was developed soon after S4 was released and is arguably one of the best case-studies for how well S4 performs. Whilst S4 formalises S3 to allow for a fully OO interface to be developed, its dependence on inheritance forces design decisions that quickly become problematic. This is seen in the large inheritance trees in \pkg{distr} in which one implemented distribution can be nested five child classes deep. This is compounded by the fact that S4 does not use pointer objects but instead nests objects internally. Therefore \pkg{distr} has problems with composite distributions in that they quickly become very large in size, for example a mixture of two distributions can easily be around 0.5Mb, which is relatively large. In contrast, R6 introduces pointers, which means that a wrapped object simply points to its wrapped component and does not copy it needlessly.

Whilst a fully object-oriented interface can be developed in S3 and S4, they do not have the flexibility of R6, which means that in the long run, extensibility and scalability can be problematic. R6 forces R users to learn a paradigm that they may not be familiar with but packages like \pkg{R62S3} allow users to become acquainted with R6 on a slightly shallower learning curve. Speed differences for the three paradigms are formally compared on the example above using \pkg{microbenchmark} \citep{pkgmicrobm}, the results are in table \ref{tab:speed}. The R6 example is compared both including construction of the class, \code{duck\$new("Zaphod")\$quack()}, and without construction, \code{d\$quack()}, where \code{d} is the object constructed before comparison. A significant `bottleneck' is noted when construction is included in the comparison but despite this S4 is still significantly the slowest.

\begin{table}[h]
\centering
\begin{tabular}{ccc}
\toprule
Paradigm & mean $(\mu s)$ & cld \\
\hline
S3 & 73.44 & a \\
S4 & 276.17 & c \\
R6 & 187.70 & b \\
R6* & 38.32 & a \\
\bottomrule
\end{tabular}
\caption{Comparing S3, S4, and R6 in calling a method. R6 is tested both including object construction (R6) and without (R6*). `cld' is the significance testing from \pkg{microbenchmark} where `a' is the fastest and `c' the slowest. Experiment conducted with \pkg{R6} version 2.4.1, \pkg{microbenchmark} version 1.4.7, and R version 4.0.2 (2020-06-22) on platform: x86\_64-apple-darwin17.0 (64-bit) running under: macOS Catalina 10.15.3.}
\label{tab:speed}
\end{table}

\subsection{Design patterns}
\label{sec:tools_distr6_patterns_intro}

In the simplest definition, `design patterns' are abstract solutions to common coding problems. They are probably most widely known due to the book `Design Patterns Elements of Reusable Object-Oriented Software' (\textit{Design Patterns}) \citep{Gamma1996}. \pkg{distr6} primarily makes use of the following design patterns
\begin{itemize}
\item Abstract Factory
\item Decorator
\item Composite
\item Strategy
\end{itemize}

\paragraph{Strategy}
The strategy pattern is common in modelling toolboxes, in which multiple algorithms can be used to solve a problem. This pattern defines an abstract class for a given problem and concrete classes that each implement different strategies, or algorithms, to solve the problem. For example in the context of mathematical integration (a common problem in R), one could use Simpson's rule, Kronrod's, or many others. These can be specified by an \code{integrate} abstract class with concrete sub-classes \code{simpson} and \code{kronrod}.

\begin{figure}[h]
\centering
\begin{tikzpicture}
\umlsimpleclass[type = abstract]{integrate}
\umlsimpleclass[x = -2, y = -1]{simpson}
\umlsimpleclass[x = 2, y = -1]{kronrod}
\umlinherit[geometry=-|]{simpson}{integrate}
\umlinherit[geometry=-|]{kronrod}{integrate}
\end{tikzpicture}
\end{figure}

\paragraph{Composite}
The composite pattern defines a collection of classes that have an identical interface when treated independently or when composed into a single class with constituent parts. To the user, this means that only one interface needs to be learnt in order to interact with composite or individual classes. A well-built composite pattern allows users to construct complex classes with several layers of composition, and yet still be able to make use of a single interface. By inheriting from a parent class, each class and composite share a common interface. Composition is a powerful design principle that allows both modification of existing classes and reduction of multiple classes \citep{Kiraly2021}. 


\paragraph{Decorator}
Decorators add additional responsibilities to an object without making any other changes to the interface. An object that has been decorated will be identical to its un-decorated counter-part except with additional methods. This provides a useful alternative to inheritance. Whereas inheritance can lead to large tree structures in which each sub-class inherits from the previous and contains all previous methods, decorators allow the user to pick and choose with responsibilities to add. The figure below demonstrates how this is useful in a shopping cart example. The top of the figure demonstrates using inheritance, in which each sub-class adds methods to the \code{Cart} parent class. By the \code{Tax} child class there are a total of five methods in the interface. In the bottom of the figure, the decorator pattern demonstrates how the functionality for adding items and tax is separated and can be added separately.

\begin{figure}[h]
\centering
\begin{tikzpicture}
\umlclass[x = -6, y = 0]{Cart}{}{Total()}
\umlclass[x = -2, y = 0]{Add}{}{AddFruit() \\
AddVegetables()}
\umlclass[x = 3, y = 0]{Tax}{}{CalculateVAT() \\
addVAT()}
\umlinherit{Add}{Cart}
\umlinherit{Tax}{Add}

\umlclass[x = -6, y = -4.5]{Cart}{}{Total()}
\umlsimpleclass[x = -2, y = -4.5, type = abstract]{AbstractDecorator}
\umlclass[x = 3, y =-3]{AddDecorator}{}{AddFruit() \\
AddVegetables()}
\umlclass[x = 3, y = -6]{TaxDecorator}{}{CalculateVAT() \\
addVAT()}
\umlinherit{AbstractDecorator}{Cart}
\umlinherit[geometry=-|]{AddDecorator}{AbstractDecorator}
\umlinherit[geometry=-|]{TaxDecorator}{AbstractDecorator}
\end{tikzpicture}
\end{figure}

\subsection{Contributions to R6}
\label{sec:tools_distr6_oop_cont}

In order to implement \pkg{distr6}, several contributions were made to the R6 paradigm, to extend its abilities and to implement the design patterns discussed above.

\paragraph{Abstract classes}
R6 did not have a concept of abstract classes, which meant that patterns such as adapters, composites, and decorators, could not be directly implemented without problems. This is produced in \pkg{distr6} with the \code{abstract} function, which is placed in the first line of all abstract classes. In the example below, \code{obj} expects the \code{self} argument from R6 classes, and \code{class} is the name of the class, \code{getR6Class} is a custom function for returning the name of the class of the given object.

\begin{verbatim}
abstract <- function(obj, class) {
  if (getR6Class(obj) == class) {
      stop(sprintf("%s is an abstract class that can't be initialized.", class))
    }
}
\end{verbatim}

For example in decorators the following line is placed at the top of the \code{initialize} function:

\begin{verbatim}
abstract(self, "DistributionDecorator")
\end{verbatim}

\paragraph{Decorators}
The typical implementation of decorators is to have an abstract decorator class with concrete decorators inheriting from this, each with their own added responsibilities. In \pkg{distr6} this is made possible by defining the \code{DistributionDecorator} abstract class (see above) with a public \code{decorate} method. Concrete decorators are simply R6 classes where the public methods are the ones to `copy' to the decorated object.

\begin{verbatim}
> DistributionDecorator
<DistributionDecorator> object generator
  Public:
    packages: NULL
    initialize: function ()
    decorate: function (distribution, ...)
    clone: function (deep = FALSE)

> CoreStatistics
<CoreStatistics> object generator
  Inherits from: <DistributionDecorator>
  Public:
    mgf: function (t)
    cf: function (t)
    pgf: function (z)
\end{verbatim}

When the \code{\$decorate} method from a constructed decorator object is called, the methods are simply copied from the decorator environment to the object environment. The \code{decorator()} function simplifies this for the user.
%
%

%

\paragraph{Composite and wrappers}
The composite pattern is made use of in what \pkg{distr6} calls `wrappers'. Again this is implemented via an abstract class (\code{DistributionWrapper}) with concrete sub-classes.

\begin{verbatim}
> DistributionWrapper
<DistributionWrapper> object generator
  Inherits from: <Distribution>
  Public:
    initialize: function (distlist = NULL, name, short_name, description, support,
    wrappedModels: function (model = NULL)
    setParameterValue: function (..., lst = NULL, error = "warn")
  Private:
    .wrappedModels: list

> TruncatedDistribution
<TruncatedDistribution> object generator
  Inherits from: <DistributionWrapper>
  Public:
    initialize: function (distribution, lower = NULL, upper = NULL)
    setParameterValue: function (..., lst = NULL, error = "warn")
  Private:
    .pdf: function (x, log = FALSE)
    .cdf: function (x, lower.tail = TRUE, log.p = FALSE)
    .quantile: function (p, lower.tail = TRUE, log.p = FALSE)
    .rand: function (n)
\end{verbatim}

Wrappers in \pkg{distr6} alter objects by modifying either their public or private methods. Therefore an `unwrapped' distribution looks identical to a `wrapped' one, despite inheriting from different classes. This is possible via two key implementation strategies: 1) on construction of a wrapper, parameters are prefixed with a unique ID, meaning that all parameters can be accessed at any time; 2) the \code{wrappedModels} public field allows access to the original wrapped distributions. These two factors allow any new method to be called either by reference to \code{wrappedModels} or by using \code{\$getParameterValue} with the newly prefixed parameter ID. This is demonstrated in the \code{.pdf} private method of the \code{TruncatedDistribution} wrapper (slightly abridged):

\begin{verbatim}
.pdf = function(x, log = FALSE) {
      dist <- self$wrappedModels()[[1]]
      lower <- self$getParameterValue("trunc_lower")
      upper <- self$getParameterValue("trunc_upper")

        pdf <- numeric(length(x))
        pdf[x > lower & x <= upper] <- dist$pdf(x[x > lower & x <= upper]) /
          (dist$cdf(upper) - dist$cdf(lower))

      return(pdf)
    }
\end{verbatim}

As the public \code{pdf} is the same for all distributions, and this is inherited by wrappers, only the private \code{.pdf} method needs to be altered.

\section{Examples}

This final section looks at concrete short examples for four key use-cases.

\subsection{Constructing and querying distributions}

The primary use-case for the majority of users will be in constructing distributions in order to query their results and visualise their shape.
\\\\
Below a distribution (Binomial) is constructed and queried for its distribution-specific traits and parameterisation-specific properties.

\begin{verbatim}
> b <- Binomial$new(prob = 0.1, size = 5)
> b$setParameterValue(size = 6)
> b$getParameterValue("size")
> b$parameters()
> b$properties
> b$traits
\end{verbatim}

Specific methods from the distribution are queried as well.

\begin{verbatim}
> b$mean()
> b$entropy()
> b$skewness()
> b$kurtosis()
> b$cdf(1:5)
\end{verbatim}

\begin{figure}[h]
\centering
\includegraphics[width = 14cm, height = 9cm]{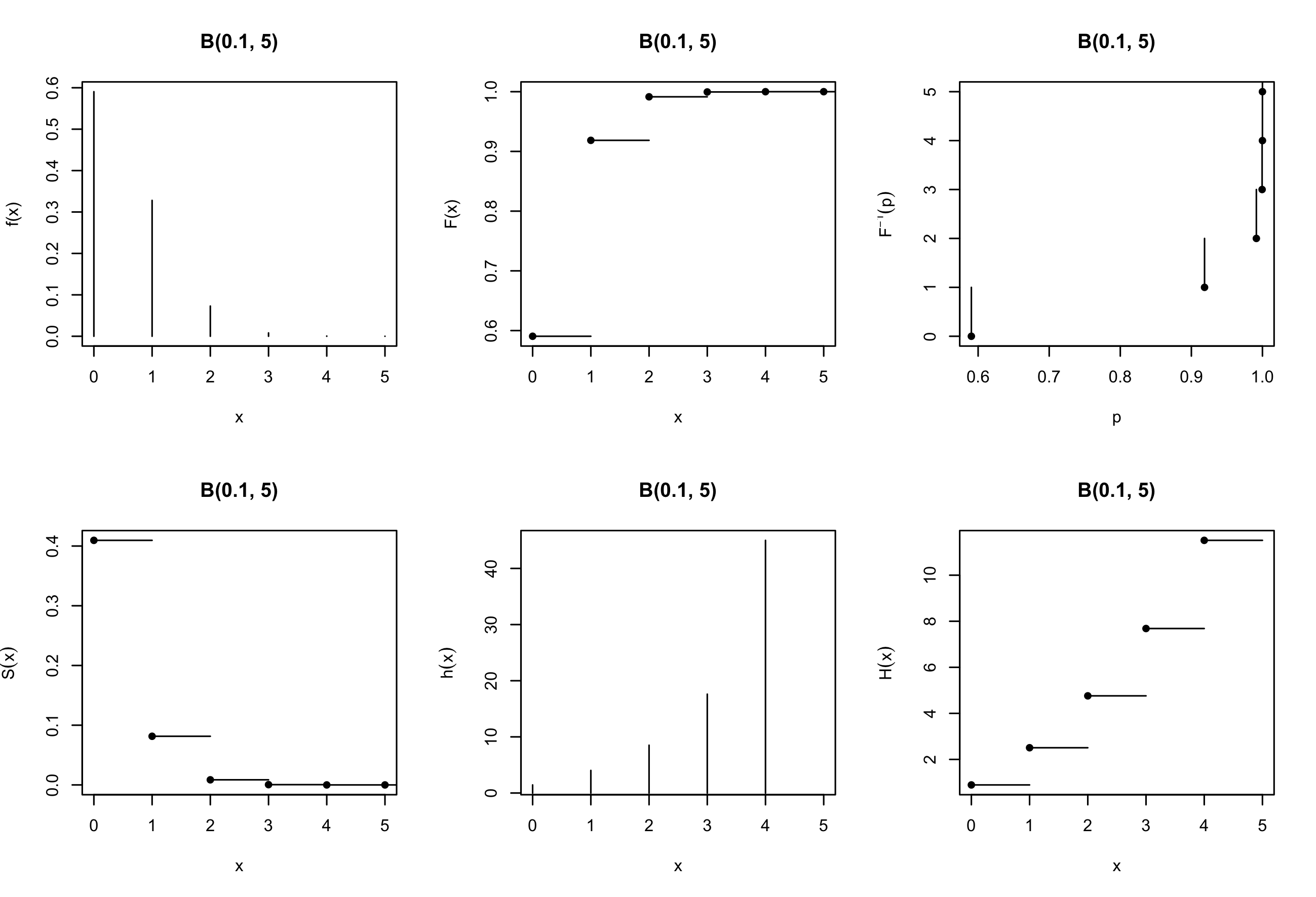}
\caption{Visualising a Binomial(0.1, 5) distribution, \code{b}, with \code{plot(b, fun = "all")}. Functions clockwise from top-left: probability mass, cumulative distribution, quantile (inverse cdf), cumulative hazard, hazard, and survival.}
\label{fig:binom}
\end{figure}

The distribution is visualised by plotting it's density, distribution, inverse distribution, hazard, cumulative hazard, and survival function; the output is in figure \ref{fig:binom}.

\begin{verbatim}
> plot(b, fun = "all")
\end{verbatim}

\subsection{Analysis of empirical data}

\pkg{distr6} can also serve as a toolbox for analysis of empirical data by making use of the three `empirical' distributions: \code{Empirical}, \code{EmpricalMV}, and \code{WeightedDiscrete}.
\\\\
First an empirical distribution is constructed with samples from a standard exponential distribution.

\begin{verbatim}
> E <- Empirical$new(samples = rexp(10000))
\end{verbatim}

The \code{summary} function is used to quickly obtain key information about the empirical distribution.

\begin{verbatim}
> summary(E)

Empirical Probability Distribution.

  Quick Statistics
	    Mean:		0.105954
	    Variance:	1.140673
	    Skewness:	0.05808027
	    Ex. Kurtosis:	-0.473978

 Support: (-2.50, -2.19,...,2.27, 2.66) 	Scientific Type: R

 Traits:	discrete; univariate
 Properties:	asymmetric; platykurtic; positive skew
\end{verbatim}

\begin{figure}[h]
\centering
\includegraphics[width = 13cm, height = 6.5cm]{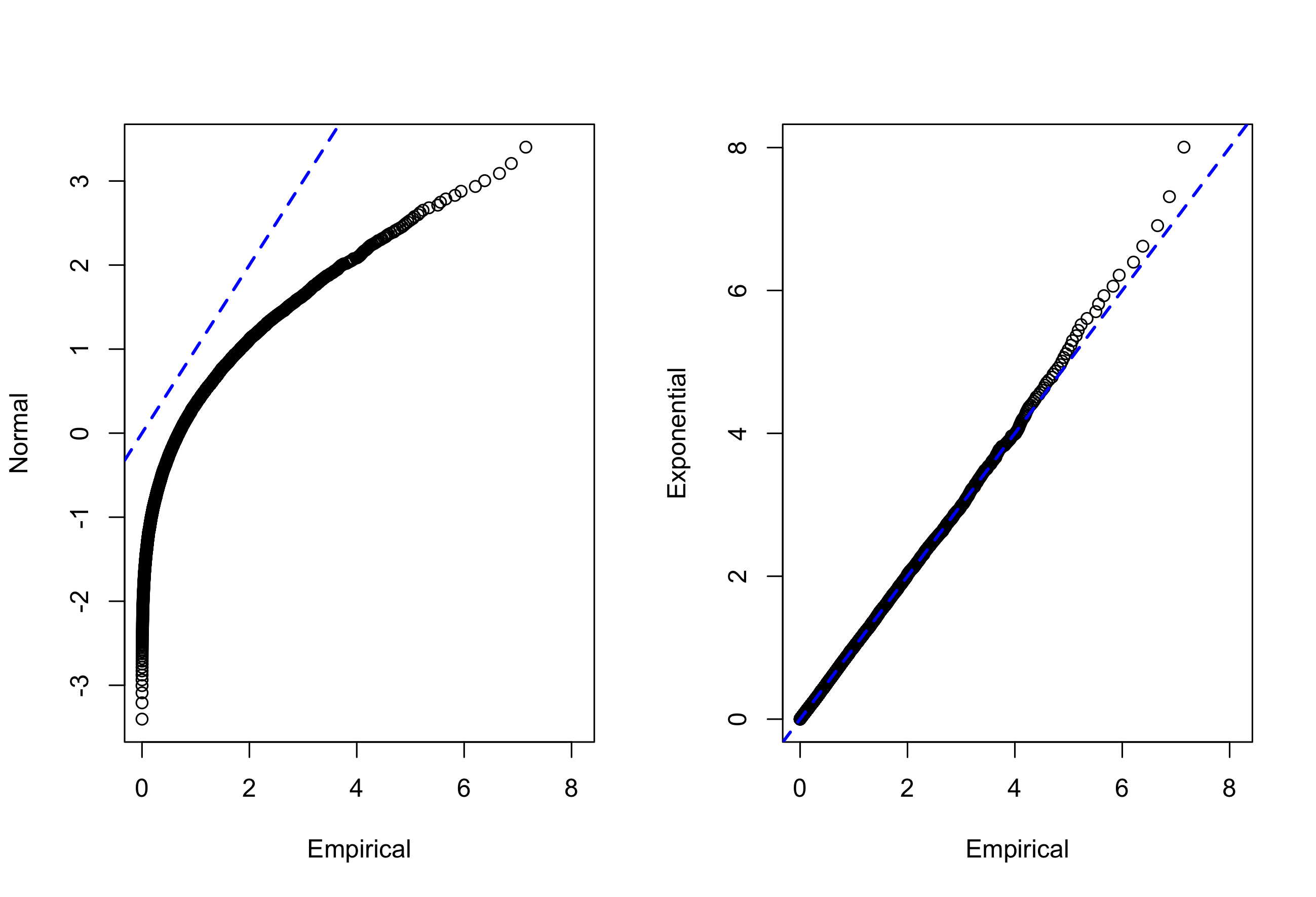}
\caption{Q-Q plots comparing an unknown Empirical distribution, \code{E}, to theoretical Normal (left) and Exponential (right) distributions with \code{qqplot(E, Normal\$new()} and \code{qqplot(E, Exponential\$new()} respectively.}
\label{fig:qqplot}
\end{figure}

The distribution is compared to a (standard) Normal distribution and then (standard) Exponential distribution; output in figure \ref{fig:qqplot}.

\begin{verbatim}
> qqplot(E, Normal$new(), xlab = "Empirical", ylab = "Normal")
> qqplot(E, Exponential$new(), xlab = "Empirical", ylab = "Exponential")
\end{verbatim}

\begin{figure}[h]
\centering
\includegraphics[width = 9cm, height = 9cm]{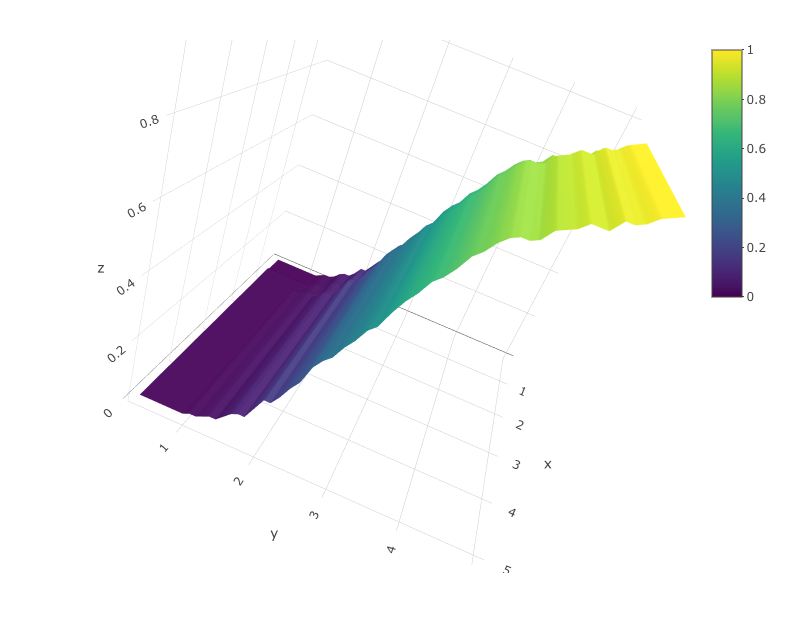}
\caption{Visualising a bivariate empirical distribution, \code{E}, with \code{plot("E", fun = "cdf")}.}
\label{fig:empmv}
\end{figure}

The CDF of a bivariate empirical distribution is visualised; output in figure \ref{fig:empmv}.

\begin{verbatim}
> plot(EmpiricalMV$new(data.frame(rnorm(100, mean = 3), rnorm(100))), fun = "cdf")
\end{verbatim}

\subsection{Learning from custom distributions}

Whilst empirical distributions are useful when data samples have been generated, custom distributions can be used to build an entirely new probability distribution - though here we use a simple discrete uniform distribution. This example highlights the power of decorators to estimate distribution results without manual computation of every possible method. The output demonstrates the precision and accuracy of these results.
\\\\
Below, a custom distribution is created and by including the \code{decorators} argument, all further methods are imputed numerically.  The distribution is summarised for properties, traits and common results (this is possible with the `CoreStatistics' decorator). The summary is identical to the analytic \code{DiscreteUniform} distribution.

\begin{verbatim}
> U <- Distribution$new(
+         name = "Discrete Uniform",
+         type = set6::Integers$new(), support = set6::Set$new(1:10),
+         pdf = function(x) ifelse(x < 1 | x > 10, 0, rep(1/10,length(x))),
+         decorators = c("CoreStatistics", "ExoticStatistics", "FunctionImputation"))
> summary(U)

Discrete Uniform

  Quick Statistics
	    Mean:		5.5
	    Variance:	8.25
	    Skewness:	0
	    Ex. Kurtosis:	-1.224242

 Support: {1, 2,...,9, 10} 	Scientific Type: Z

 Traits:	discrete; univariate
 Properties:	asymmetric; platykurtic; no skew

 Decorated with:  CoreStatistics, ExoticStatistics, FunctionImputation
\end{verbatim}

The CDF and simulation function are called (numerically imputed with the \code{FunctionImputation} decorator), the hazard function from the \code{ExoticStatistics} decorator, and the \code{kthmoment} function from the \code{CoreStatistics} decorator.
\begin{verbatim}
> U$cdf(1:10)
[1] 0.1 0.2 0.3 0.4 0.5 0.6 0.7 0.8 0.9 1.0
> U$rand(10)
[1]  8 10  5  8  5 10  6  7  1  4
> U$hazard(2)
[1] 0.125
> U$kthmoment(2)
[1] 8.25
\end{verbatim}

\subsection{Composite distribution modelling}

Composite distributions are an essential part of any distribution software, the following example demonstrates two types of composites: composition via distribution transformation (truncation), and composition via mixtures and vectors.
\\\\
First, a Binomial distribution is constructed and truncated between $1$ and $5$, the CDF of the new distribution is queried.

\begin{verbatim}
> TB <- truncate(
   Binomial$new(size = 20, prob = 0.5),
   lower = 1,
   upper = 5
   )
> round(TB$cdf(0:6), 4)
[1] 0.0000 0.0000 0.0088 0.0613 0.2848 1.0000 1.0000
\end{verbatim}

Next, a vector distribution is constructed of two Normal distributions, with respective means $1$ and $2$ and unit standard deviation. The parameters are queried (some columns suppressed).

\begin{verbatim}
> V <- VectorDistribution$new(distribution = "Normal",
+                             params = data.frame(mean = 1:2))
> V$parameters()
           id value support
1: Norm1_mean     1       R
2:  Norm1_var     1      R+
3:   Norm1_sd     1      R+
4: Norm1_prec     1      R+
5: Norm2_mean     2       R
6:  Norm2_var     1      R+
7:   Norm2_sd     1      R+
8: Norm2_prec     1      R+
\end{verbatim}

Vectorisation is possible across distributions, samples, and both. In the example below, the first call to \code{\$pdf} evaluates both distributions at (1, 2), the second call evaluates the first at (1) and the second at (2), and the third call evaluates the first at (1, 2) and the second at (3, 4).

\begin{verbatim}
> V$pdf(1:2)
       Norm1     Norm2
1: 0.3989423 0.2419707
2: 0.2419707 0.3989423
> V$pdf(1, 2)
       Norm1     Norm2
1: 0.3989423 0.3989423
> V$pdf(1:2, 3:4)
       Norm1      Norm2
1: 0.3989423 0.24197072
2: 0.2419707 0.05399097
\end{verbatim}

Finally a mixture distribution with uniform weights is constructed from a $Normal(2, 1)$ distribution and an $Exponential(1)$.

\begin{verbatim}
> MD <- MixtureDistribution$new(
+   list(Normal$new(mean = 2, sd = 1),
+        Exponential$new(rate = 1)
+   )
+ )
> MD$pdf(1:5)
[1] 0.304925083 0.267138782 0.145878896 0.036153303 0.005584898
> MD$cdf(1:5)
[1] 0.3953879 0.6823324 0.8957788 0.9794671 0.9959561
> MD$rand(5)
[1] 3.6664473 0.1055126 0.6092939 0.8880799 3.4517465
\end{verbatim}

\section{Future work}
Whilst \pkg{distr6} fulfils its primary purpose as an R6 interface for probability distributions with basic features, it is not consider 'feature-complete' as it currently lacks many of the important features included in \pkg{distr} and other related software. \pkg{distr6} is in constant development and has an active \href{https://github.com/alan-turing-institute/distr6}{GitHub} with open issues and projects. Some concrete short-term goals include:
\begin{itemize}
\itemsep-0.2em
\item A more generalised convolution interface
\item Expanding truncation to other interval types (currently only left-open is supported)
\item Extending the \code{FunctionImputation} decorator to work on higher-order distributions as well as to improve speed and accuracy.
\item Exposing further internal functionality for more user-control over numerical results.
\end{itemize}

\section{Summary}
\pkg{distr6} introduces a robust and scalable object-oriented interface for probability distributions to R and aims to be the first-stop for class object-oriented probability distributions in R. By making use of R6, every implemented distribution is clearly defined with properties, traits, and analytic results. Whilst R \pkg{stats} is limited to very basic \code{dpqr} functions for representing evaluated distributions, \pkg{distr6} ensures that probability distributions are treated as complex mathematical objects.

Future updates of the package will include adding further numerical approximation strategies in the decorators to allow users to choose different methods (instead of being forced to use one). Additionally, the extensions to R6 could be abstracted into an independent package in order to better benefit the R community.

\pkg{distr6} is released under the MIT licence on \href{https://github.com/alan-turing-institute/distr6}{GitHub} and \href{https://CRAN.R-project.org/package=distr6}{CRAN}. Extended documentation, tutorials, and examples are available at \url{https://alan-turing-institute.github.io/distr6/}. Code quality is monitored and maintained by an extensive suite of unit tests on multiple continuous integration systems.


\section*{Acknowledgments}

We would like to thank and acknowledge Prof. Dr. Peter Ruckdeschel and Prof. Dr. Matthias Kohl for their work on the \pkg{distr} package and for extensive discussions, planning, and design decisions that were utilised in the development of \pkg{distr6}. RS receives a PhD stipend from EPSRC (EP/R513143/1).

\newpage
\bibliography{distr6}

\end{document}